\documentclass[fleqn,usenatbib]{mnras}

\usepackage[T1]{fontenc}
\usepackage{ae,aecompl}

\usepackage{graphicx}	% Including figure files
\usepackage{amsmath}	% Advanced maths commands
\usepackage{pdflscape}
\usepackage{afterpage}
\usepackage{rotating}
\usepackage{newtxtext,newtxmath}

\newcommand{\xmm}{{\it XMM-Newton}}

\newcommand{\red}{\textcolor{black}}%{}%

\newcommand{\blue}{\textcolor{black}}
\newcommand{\third}{\textcolor{black}}
\newcommand{\srca}{RBS~2041}
\newcommand{\srcb}{RX~J0439}
\newcommand{\srcc}{RX~J0136}
\newcommand{\srcd}{RX~J1355}
\newcommand{\srce}{1ES~0919}
\newcommand{\srcafull}{RBS~2041}
\newcommand{\srcbfull}{RX~J0439.6$-$5311}
\newcommand{\srccfull}{RX~J0136.9$-$3510}
\newcommand{\srcdfull}{RX~J1355.2$+$5612}
\newcommand{\srcefull}{1ES~0919$+$515}

\title[Ultra-Soft NLS1s]{A Disc Reflection Model for Ultra-Soft Narrow-Line Seyfert 1 Galaxies}

\author[J. Jiang et al.]{Jiachen Jiang,$^{1,2,3}$\thanks{E-mail: jcjiang@tsinghua.edu.cn}
Luigi C. Gallo,$^{4}$
Andrew C. Fabian,$^{3}$
Michael L. Parker$^{5}$
\newauthor
and Christopher S. Reynolds$^{3}$\\
$^{1}$Tsinghua Center for Astrophysics, Tsinghua Univerisity, Shuangqing Road, Beijing 100084, China\\
$^{2}$Department of Astronomy, Tsinghua Univerisity, Shuangqing Road, Beijing 100084, China\\
$^{3}$Institute of Astronomy, University of Cambridge, Madingley Road, Cambridge CB3 0HA, UK\\
$^{4}$Department of Astronomy and Physics, Saint Mary's University, 923 Robie Street, Halifax, NS, B3H 3C3, Canada \\
$^{5}$European Space Agency, European Space Astronomy Centre, E-28691 Villanueva de la Ca\~nada, Spain\\
}

\date{Accepted XXX. Received YYY; in original form ZZZ}

\pubyear{2019}

\begin{document}
\label{firstpage}
\pagerange{\pageref{firstpage}--\pageref{lastpage}}
\maketitle

\begin{abstract}
We present a detailed analysis of the \xmm\ observations of five narrow-line Seyfert 1 galaxies (NLS1s). They all show very soft continuum emission in the X-ray band with a photon index of $\Gamma\gtrsim 2.5$. Therefore, they are referred to as `ultra-soft' NLS1s in this paper. By modelling their optical/UV--X-ray spectral energy distribution (SED) with a reflection-based model, we find \blue{indications} that the disc surface in these ultra-soft NLS1s is in a higher ionisation state than other typical Seyfert 1 AGN. Our best-fit SED models suggest that these five ultra-soft NLS1s have an Eddington ratio of $\lambda_{\rm Edd}=1-20$ assuming available black hole mass measurements. In addition, our models infer that a significant fraction of the disc energy in these ultra-soft NLS1s is radiated away in the form of non-thermal emission instead of the thermal emission from the disc. \red{Due to their extreme properties, X-ray observations of these sources in the iron band are particularly challenging. Future observations, e.g. from \textit{Athena}, will enable us to have a clearer view of the spectral shape in the iron band and thus distinguish the reflection model from other interpretations of their broad band spectra.} 
\end{abstract}

\begin{keywords}
accretion, accretion discs\,-\,black hole physics, X-ray: galaxies, galaxies: Seyfert
\end{keywords}

%%%%%%%%%%%%%%%%%%%%%%%%%%%%%%%%%%%%%%%%%%%%%%%%%%

%%%%%%%%%%%%%%%%% BODY OF PAPER %%%%%%%%%%%%%%%%%%

\section{Introduction}

Narrow-line Seyfert 1 galaxies (NLS1s) are a unique class of Seyfert 1 galaxies (Sy1s). They are similar to other Sy1s, except for having strong  Fe \textsc{ii} emission, weak [O \textsc{iii}] emission, and a narrow H$\beta$ line \citep[e.g.][]{osterbrock77,goodrick89}. According to the definition of NLS1, the full-width at half maxium (FWHM) of their H$\beta$ lines is smaller than 5000\,km\,s$^{-1}$ \citep{goodrick89}. These narrow H$\beta$ lines are believed to be related to the small black hole (BH) masses in NLS1s, assuming H$\beta$ emission is related to the broad-line region \citep[e.g. BLR,][]{grupe04}. However, \citet{marconi08} points out that NLS1s and other Sy1s may have similar BH masses if the radiation pressure onto the BLR is taken into account. This is particularly important in NLS1s, where the disc is often found to have a near-Eddington accretion rate. 

%There has been increasing evidence for radio-loud and gamma-ray emitting NLS1s. These NLS1s are similar to blazars but have much weaker jet power \cite[e.g.][]{abdo09,paliya19}. The evidence of radio and gamma-ray emission suggests that NLS1s are also able to launch jet despite their moderate BH masses \citep[e.g.][]{yuan08}.

In the X-ray band, NLS1s often show unique properties, such as very soft continuum emission and highly variable soft excess emission \citep[e.g.][]{boller96,gallo18}. \citet{gallo06} classifies NLS1s into two general categories according to the variability of their optical and X-ray emission: `complex' and `simple' NLS1s. `Complex' NLS1s often show larger X-ray flux variability than the `simple' ones. For example, 1H~0707$-$495 and IRAS~13224$-$3809, classified as `complex' NLS1s, show very fast and large flux variability on kilosecond timescales \citep[e.g.][]{boller03, fabian04, alston19}. The X-ray complexity of these NLS1s is often explained by either the light-bending model in the reflection scenario \citep[e.g.][]{miniutti06, jiang19}, or variable ionised absorption fully or partially covering the central emission region \citep[e.g.][]{done16}. It is important to mention the increasing number of discoveries of X-ray reverberation lags. They are seen in the soft X-ray band \citep[e.g.][]{fabian09,demarco13}, the iron band \citep[e.g.][]{kara16}, and the hard X-ray band \citep[e.g.][]{zoghbi14,kara15} of some AGN that have no obvious evidence of strong ionised absorption features in their spectra. In addition, detailed principle component analysis also shows that the X-ray variability agrees with the reflection scenario in these unabsorbed sources \citep{parker14,parker15b}. Similar X-ray reverberation lags and spectral properties have also been seen in BH X-ray binaries \citep[e.g.][]{reis13, demarco15, kara19, mastroserio19}.

In this work, we present detailed spectral analysis for five extreme NLS1s: \srcafull, \srcbfull, \srccfull, \srcdfull\, and \srcefull. See Table\,\ref{tab_obs} for further information about them. These sources have been identified in the \textit{ROSAT} soft X-ray survey \citep{voges99}. They show extremely soft emission in the soft X-ray band, and potentially host a BH that is accreting around or above the Eddington limit. Due to the extreme steepness of their X-ray spectra, we refer them as ultra-soft NLS1s in this paper. The high accretion rates of these ultra-soft NLS1s are particularly interesting. For instance, we might be able to understand the existence of massive quasars in the early universe by studying these nearby sources \citep[e.g.][]{wu15,banados16,banados18,tang19}.  

Previously, the soft X-ray emission from these ultra-soft NLS1s was modelled by warm corona models, where there is an optically-thick corona in addition to the optically-thin hot corona \citep[e.g.][]{jin09,jin17}. The temperature of this extra corona is usually below 1~keV \red{\citep{jin12,petrucci18,panda19}}, which is much lower than the hot corona. Therefore, they are often referred to as the `warm' corona. In this scenario, the UV emission from ultra-soft NLS1s was often found to be dominated by this warm coronal emission instead of the thermal emission from the disc, and the discs were found to accrete at a super-Eddington accretion rate of more than 10 times the Eddington limit \citep[e.g.][]{jin09}. 

An alternative explanation of soft excess emission is the reflection from the innermost region of the accretion disc \citep[e.g.][]{crummy06, walton13, jiang19c}. In this scenario, the disc is illuminated by the non-thermal emission from the hot corona, and produces reprocess spectrum \red{within the Thomson optical depth of the disc}. The reprocess spectrum is referred to as the disc `reflection' spectrum, which consists of series of emission lines in the soft X-ray band and a Compton back-scattered continuum in the hard X-ray band \citep[e.g.][]{ross93,garcia10}. The emission lines are broadened by strong relativistic effects in the vicinity of BHs \citep[e.g.][]{reynolds19}. 

In this paper, we systematically apply this relativistic disc reflection model to the \xmm\ data of five ultra-soft NLS1s, and study their broad-band spectral energy distribution (SED) based on our reflection modelling. In Section\,\ref{data}, we introduce the data reduction process. In Section\,\ref{xray}, we analyse the X-ray spectra of these sources by using disc reflection model. We also show supporting Markov Chain Monte Carlo (MCMC) analysis of the X-ray data in Appendix\,\ref{mcmc}. In Section\,\ref{sed}, we model their SEDs by extending our reflection model to the optical and UV bands. 
%In Section\,\ref{jin}, we consider a warm corona interpretation of the soft excess emission by using \srccfull\ as an example, and point out the limitations and inconsistency in the current warm corona for this source. 
In Section\,\ref{conclude}, we summarise our results. 

%\citet{ai11}
%rosat catalog \citep{voges99}

\section{\xmm\ Data Reduction} \label{data}

\begin{figure*}
    \centering
    \includegraphics[width=17cm]{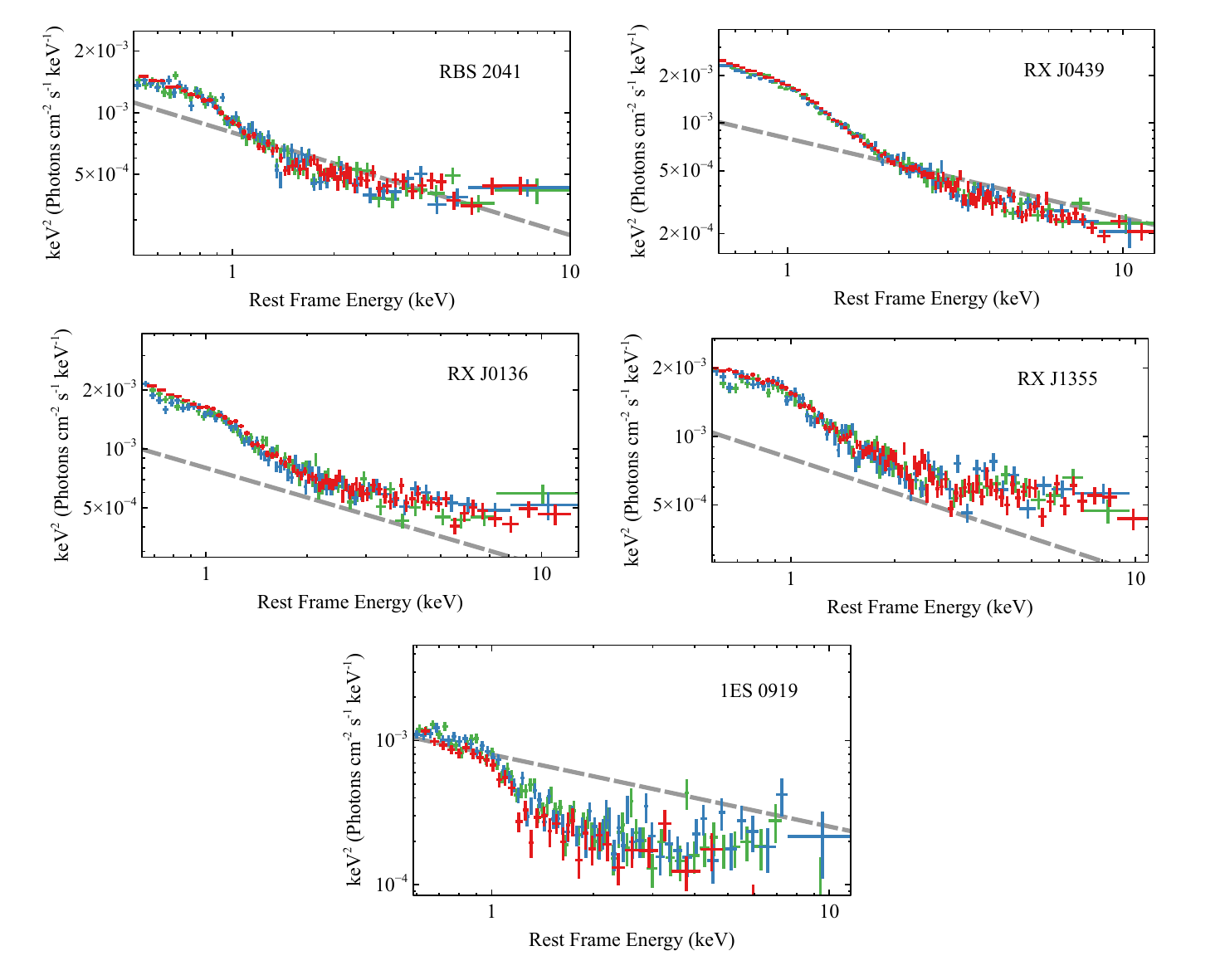}
    \caption{Unfolded spectra of the five NLS1s in our work. Red: pn; blue: MOS1; green: MOS2. A power-law model with $\Gamma=0$ is used to unfold the spectra. The dashed gray line in each panels show a power law with $\Gamma=2.5$ in comparison with the data.}
    \label{pic_eeuf}
\end{figure*}

We use the European Photon Imaging Camera (EPIC) observations for X-ray continuum modelling, and the Optical Monitor (OM) observations for flux measurements in the optical and UV bands. A full list of observations used in our work is in Table\,\ref{tab_obs}.

The EPIC data are reduced using V17.0.0 of the \xmm\ Science Analysis System (SAS) software package. The version of the calibration files  is v.20180620. We first generate a clearn event file by running EMPROC (for EPIC-MOS data) and EPPROC (for EPIC-pn data). Then, we select good time intervals by filtering out the intervals that are dominated by flaring particle background. These high-background intervals are where the single event (PATTERN=0) count rate in the >10~keV band is larger than 0.35~counts~s$^{-1}$ (0.4 counts~s$^{-1}$) for MOS (pn) data. By running the EVSELECT task, we select single and double events for EPIC-MOS (PATTERN<=12) and EPIC-pn (PATTERN<=4, FLAG==0) source event lists from a circular source region. No obvious evidence of pile-up effects has been found in our observations. The background spectra are extracted from nearby regions on the same unit. Last, we create redistribution matrix files and ancillary response files by running RMFGEN and ARFGEN. 

We consider the EPIC spectra between 0.5--10~keV. The EPICSPECCOMBINE tool is used to make a stacked spectrum for each camera, along with corresponding background spectra and response matrix files, if there are multiple observations for one source. We do not merge spectra from different instruments. The SPECGROUP command is used to group the spectra such that each bin has a minimum number of 20 counts and a minimum width that is 1/3 of the resolution at that energy.

We reduce OM data using the OMICHAIN tool. In order to convert the flux obtained by OM into the XSPEC data format, we apply the OM2PHA tool to the combined source list of each observation. The corresponding OM response files can be found on the \xmm\ website\footnote{ftp://xmm.esac.esa.int/pub/ccf/constituents/extras/responses/OM}.

\begin{table*}
    \caption{List of \xmm\ observations analysed in this work. The redshift values are from the NED website. The last column shows the net exposure of MOS1, MOS2 and pn observations respectively after removing the time intervals that are dominated by flaring particle background. \red{ Column 3: The BH masses were estimated by measuring $H_{\beta}$ line widths. Column 4: References for $M_{\rm BH}$ measurements.}}
    \label{tab_obs}
    \centering
    \begin{tabular}{cccccccccc}
    \hline\hline
    Source & Full Name & $M_{\rm BH}$ & Ref & $z$ & $N_{\rm H}$ & $E(B-V)$ & Obs ID & Net Expo\\
           & & $10^{6}M_{\odot}$ & & & $10^{20}$ cm$^{-2}$ & & & ks \\
    \hline
    \srca  & \srcafull & 10 & \citet{grupe10} &  0.137 & 2.16 & 0.029 & 0741390301 & 34, 34, 28\\

    \srcb  & \srcbfull& 3.9 &  \citet{grupe10} & 0.243 & 0.82 & 0.006 & 0741390101 & 25, 25, 18\\
           & &    &       & &      &       & 0764530101 & 131, 130, 122\\
    \srcc  & \srccfull& 79  & \citet{jin09} & 0.289 & 2.17 & 0.018 & 0303340101 & 50, 50, 38\\
    \srcd  & \srcdfull& 6.7 & \citet{grupe10} & 0.122 & 1.05 & 0.010 & 0741390201 & 23, 23, 16\\
           & &    &       &  &    &       & 0741390401 & 22, 26, 19\\
    \srce  & \srcefull& 5.0   & \citet{komossa08} & 0.159 & 1.37 & 0.014 & 0300910301 & 23, 23, 4\\
    \hline\hline
    \end{tabular}
\end{table*}

\section{X-ray Spectral Analysis} \label{xray}

We use XSPEC V12.10.1h \citep{arnaud96} for spectral analysis, and $\chi^{2}$ is considered in this work. The column density of the Galactic absorption along the line of sight towards our sources is calculated by \citet{willingale13}, which can be found in Table\,\ref{tab_obs}. The \texttt{tbnew} model \citep{wilms00} is used to account for Galactic absorption, and the \texttt{zdust} model \citep{pei92} is used to account for Galactic extinction. We fix the column density of the Galactic absorption at the values given by \citet{willingale13} during our spectral fitting as they are all very low and cannot be constrained by our data. The luminosity distances of our sources are from the NED website, where $H_0$=67.8\,km\,s$^{-1}$\,Mpc$^{-1}$, $\Omega_{\rm matter}$ = 0.308, and $\Omega_{\rm vacuum}$ = 0.692 are assumed.

\begin{figure}
    \centering
    \includegraphics[width=\columnwidth]{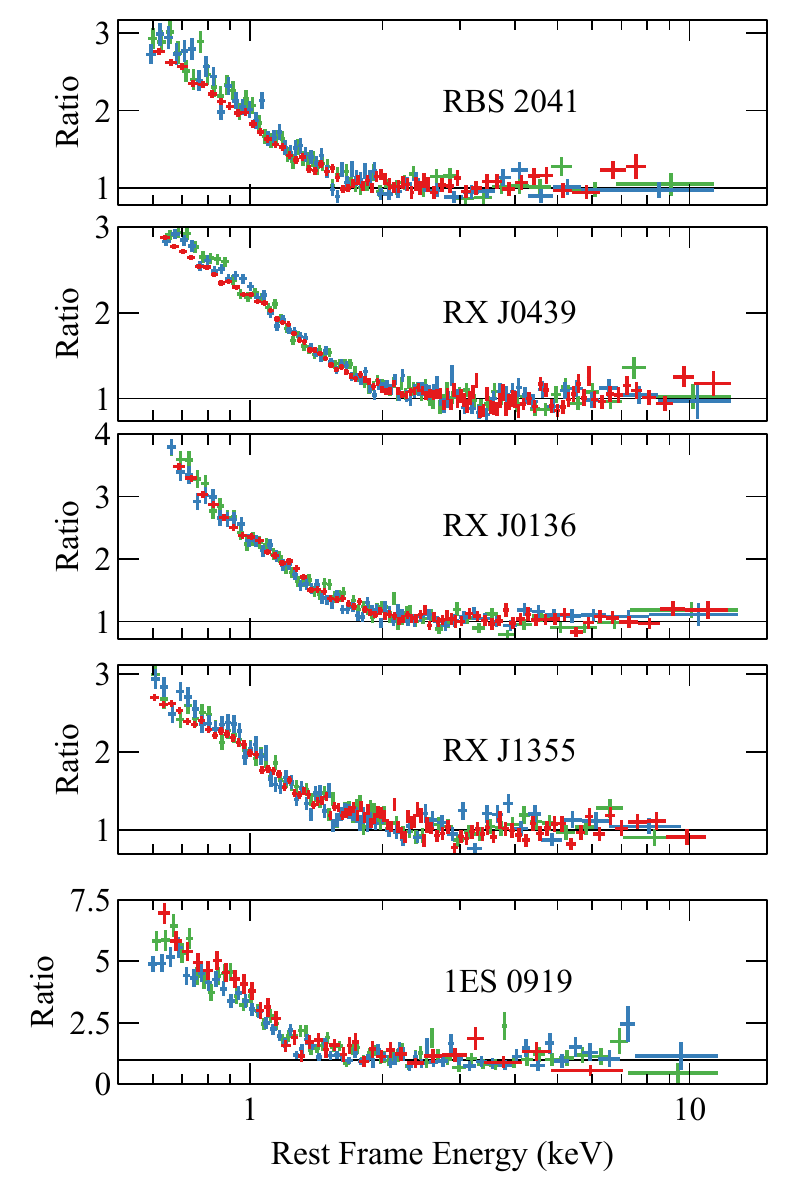}
    \caption{Data/model ratio plots using the best-fit absorbed power-law models for the spectra above 2\,keV.}
    \label{pic_pl}
\end{figure}

\subsection{Reflection Model Setup}

All the X-ray spectra analysed in our work are shown in Fig.\,\ref{pic_eeuf}. They are unfolded using a power-law model with $\Gamma=0$ to remove the impact of instrumental response. The grey line in each panel shows a power-law model with $\Gamma=2.5$ and the same normalization for comparison between the data. At the first glance, all the spectra show a continuum softer than $\Gamma=2.5$ below 3~keV, and turn harder above 3~keV. But the steepness of the spectra is slightly different in different sources. For example, the spectra of \srcb\ are consistent with $\Gamma=2.5$ above 3~keV. In comparison, the spectra of \srce\ are relatively harder, and are consistent with a power law with $\Gamma=2$ above 3~keV\footnote{A power law with $\Gamma=2$ would be a horizontal line in this figure.}. 
%In addition, \srca, \srcb, \srcc\ and \srcd\ show a similar X-ray flux while \srce\ shows a relatively lower X-ray flux level.  

We first model all the spectra above 2\,keV with an absorbed power-law model, and then include the 0.5--2\,keV band of the spectra without changing the fits. The data/model ratio plots are shown in Fig.\ref{pic_pl}. 

In the soft X-ray band, all the spectra show very steep `excess' emission below 2\,keV. The signal-to-noise in the iron band of our observations is too low due to the steepness of the intrinsic spectra and the brightness of AGN. Therefore, we are unable to determine the existence of broad Fe~K emission with high confidence. 

\red{However, it is important to mention that the lack of evidence for `apparent' broad Fe K emission line in limited-signal-to-noise (S/N) data does not rule out the existence of a reflection spectrum from the inner disc in unobscured Seyfert AGN especially when strong soft excess emission is found. The weak broad iron lines could be due to 
%Comptonisation processes of the reflection emission in a radially extended coronal region \citep[e.g.][]{wilkins15,steiner17}, 
certain disc properties \citep[e.g. a high ionisation state,][]{garcia13}, extreme relativistic effects near a spinning BH \citep[e.g.][]{crummy06} and the very soft nature of the X-ray emission. Future high S/N soft X-ray observations, e.g. from \textit{Athena}, will be able to obtain a more detailed view of these ultra-soft NLS1s in the iron band. See Section\,\ref{conclude} for simulations for future \textit{Athena} observations based on our reflection model.}

Second, we model the full-band spectra by including both disc reflection and coronal emission. The \texttt{nthcomp} model \citep{zycki99} is used to model the continuum emission from the hot corona. This model calculates the thermal Comptonisation process of cool disc seed photons in a hot coronal region. The electron temperature ($kT_{\rm e}$) of the corona decides the high-energy cutoff of the spectrum. We fix this parameter at $kT_{\rm e}=100$\,keV during our fit due to the lack of simultaneous hard X-ray data. A disc-blackbody spectrum is assumed for seed photons. The low-energy turnover, which is determined by the disc seed photon temperature $kT_{\rm db}$, is not visible in the X-ray data. Therefore, we fix this parameter at $kT_{\rm db}=10$\,eV when analysing our X-ray spectra.

An extended version of the \texttt{reflionx} model \citep{ross93} is used in our work\footnote{We do not use the \texttt{relxill} model \citep{garcia16}, which is another relativistic reflection model commonly used for spectral modelling. Because the publicly available version of \texttt{relxill} does not include the reflection spectrum below 0.1\,keV. We need a consistent model to account for the non-thermal component in the later broad-band SED modelling.}, which calculates the reprocess spectrum from an ionised slab illuminated by \texttt{nthcomp} (Jiang et al., submitted). \blue{We link the $kT_{\rm db}$, $kT_{\rm e}$, and $\Gamma$ parameters in \texttt{reflionx} to the corresponding parameters in \texttt{nthcomp}.}  Other parameters in \texttt{reflionx} are the disc iron abundance ($Z_{\rm Fe}$), the disc ionisation ($\xi$), and the density of the disc within the optical depth ($n_{\rm e}$). The \texttt{relconv} model \citep{dauser13} is applied to \texttt{reflionx} to account for relativistic correction. A phenomenological power-law disc emissivity profile parametrised by the index $q$ is used for simplicity. The other parameters in \texttt{relconv} are the disc inclination angle ($i$) and the BH spin parameter ($a_{\rm *}$). \red{The inner radius of the disc is assumed to be at the Innermost Stable Circular Orbit (ISCO).} The \texttt{constant} model is used to account for cross-calibration uncertainty between different instruments. The \texttt{cflux} model is used to calculate the flux of each component between 0.5--10keV in the observer's frame. The full model is \texttt{constant * tbnew* zdust *(cflux*relconv*reflionx + cflux*nthcomp)} in the XSPEC format. An empirical definition of reflection fraction is used here to compare the relative strength of the disc reflection component: $f_{\rm refl}=F_{\rm refl}/F_{\rm pl}$, where $F_{\rm refl}$ and $F_{\rm pl}$ are the 0.5--10\,keV band flux of the best-fit \texttt{reflionx} and \texttt{nthcomp} models. Note that this reflection fraction is different from the physical definition of reflection fraction in \citet{dauser16}.

\red{We also test for any possible narrow Fe K emission line feature from a distant cold neutral reflector by adding an additional \texttt{xillver} model. The ionisation parameter is fixed at $\log(\xi)=0$. The fits between 3--10\,keV of all of our six sources are not significantly improved. For example, \srca\ with $\Delta{\chi^{2}}=4$ and 2 more free parameters. Only an upper limit of the normalisation parameter of \texttt{xillver} is obtained (norm<$4\times10^{-6}$). Therefore, we conclude that there is no significant evidence for a distant reflector.}

\begin{figure}
    \centering
    \includegraphics[width=\columnwidth]{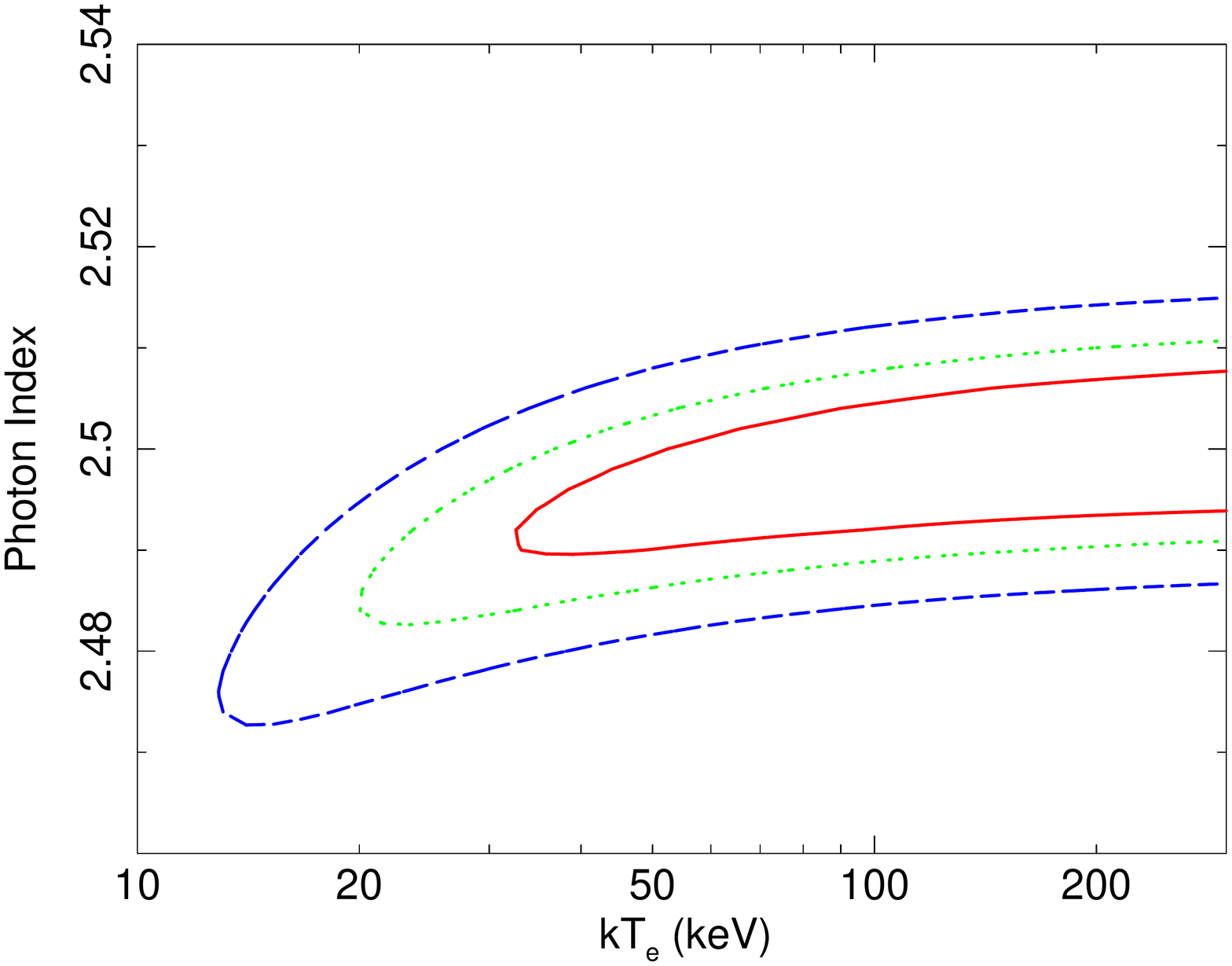}
    \caption{\red{A contour plot of $\chi^{2}$ distribution on the photon index vs. $kT_{\rm e}$ parameter plane for \srca. The lines show the $1\sigma$ (red solid line), $2\sigma$ (green dotted line), and $3\sigma$ contours (blue dashed line). Only a lower limit of the coronal temperature is obtained. See text for more details.}}
    \label{pic_ecut}
\end{figure}

\subsection{Results}

\begin{figure*}
    \centering
    \includegraphics[width=17cm]{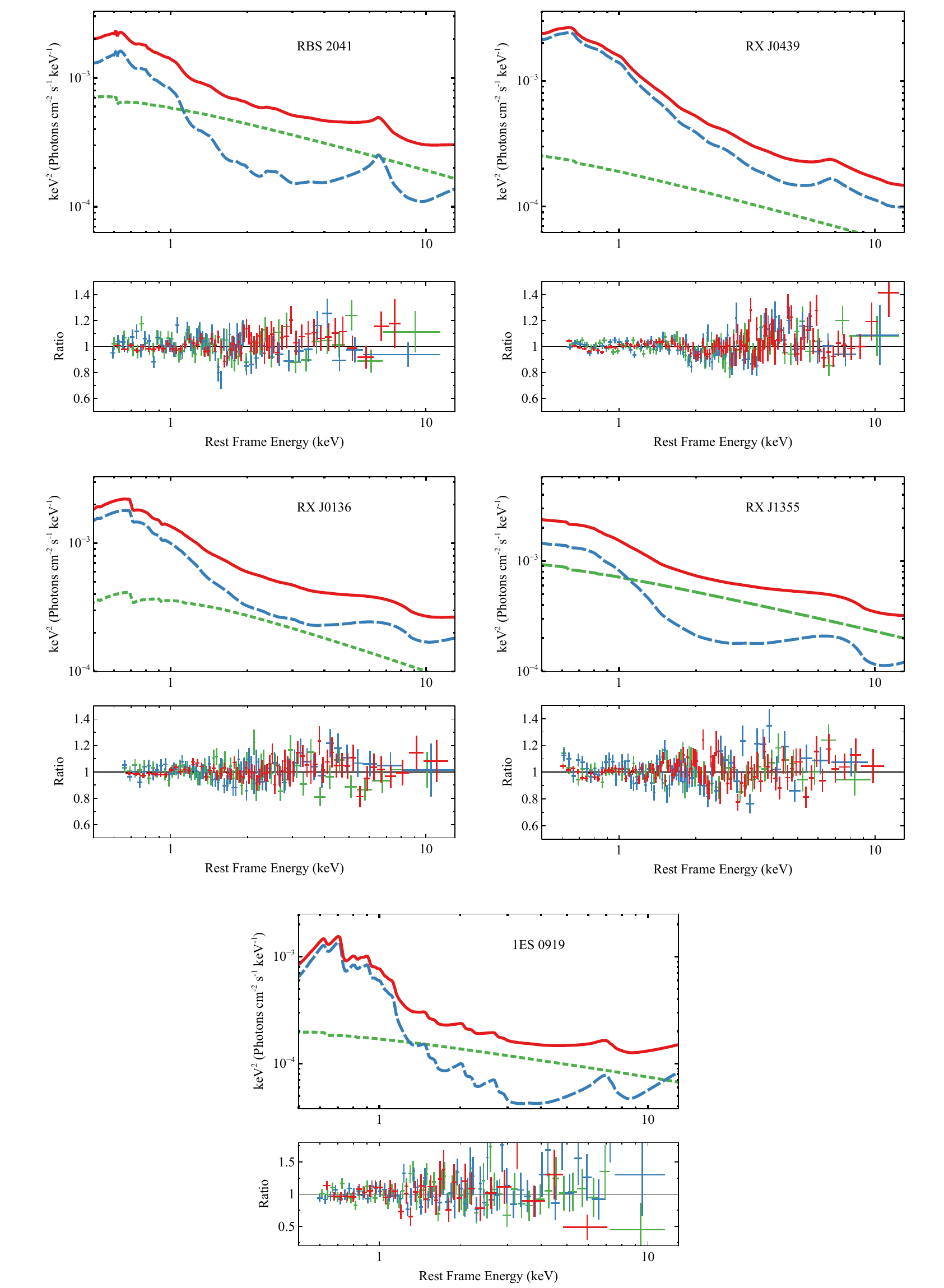}
    \caption{Best-fit reflection models and corresponding data/model ratio plots. Red solid lines: total model; blue dashed lines: relativistic disc reflection; green \red{dashed} line: Comptonisation model; red crosses: pn; blue crosses: MOS1; green crosses: MOS2.}
    \label{pic_fit}
\end{figure*}

The relativistic disc reflection model offers a very good fit for all of our X-ray spectra. The best-fit parameters can be found in Table\,\ref{tab_fit}. The best-fit models and corresponding data/model ratio plots can be found in Fig.\,\ref{pic_fit}. There are no obvious structural residuals after fitting the X-ray spectra with disc reflection model. 

We note that \srcb\ shows some excess emission above 9\,keV in the source frame (see the second panel of Fig.\,\ref{pic_fit}), which was also noticed in the previous reflection modelling in \citet{jin17}. \red{By adding a weak hard power law to account for possible jet emission as suggested in \citet{jin17}, the fit is improved above 9\,keV with $\Delta\chi^{2}=5$ and 2 more free parameters. Jet emission is considered for many other RQ NLS1 \citep[e.g.][]{wilkins17} and could be distinguished with analysis of lag-frequency spectra \citep[e.g.][]{alston20} and emissivity profiles \citep[][]{gonzalez17}.} The key parameters of the reflection model for \srcb\ do not change after adding this additional power law. Therefore, we conclude that the excess feature is statistically insignificant. Another explanation of this feature is possible calibration uncertainty near the edge of the energy range of EPIC. \red{This feature was not observed in other observations, which might be due to different configurations of the instrument when they were being operated. For instance, the EPIC-pn observation (obs ID 0764530101) of \srcb\ was the only one in our sample that was operated in the Large Window mode.} 

\red{We test whether a distant reflector without relativistic blurring is able to explain the broad band X-ray spectra. The convolution model \texttt{relconv} is removed for this test. Such a model provides a much worse fit as the model predicts narrow emission lines that are not shown in the CCD-resolution spectra. For instance, a distant reflector model offers a fit for \srca\ with $\chi^{2}/\nu=427.78/300$. The relativistic reflection model is able to offer a much better fit with $\Delta\chi^{2}=181.09$ and 3 more free parameters. Similar conclusions are found for other sources.}

\red{Additionally, we discuss the impact of the electron temperature of the corona ($kT_{\rm e}$) on our X-ray spectral modelling. $kT_{\rm e}$ determines the high energy cut-off of the X-ray spectrum. This is particularly interesting as the X-ray continuum emission is very soft in ultra-soft NLS1s. We fix this parameter at a large value ($kT_{\rm e}=100$\,keV) during the analysis above. In order to estimate how the $kT_{\rm e}$ parameter would affect our measurements of photon index, we allow this parameter to be free in the following test. For example, a $\chi^{2}$ distribution on the $\Gamma$ vs. $kT_{\rm e}$ parameter plane for \srca\ is shown in Fig.\,\ref{pic_ecut}. Due the lack of hard X-ray observations, we only obtain a lower limit of $kT_{\rm e}$. The 3-$\sigma$ lower limit is approximately 15\,keV. The photon index has a 3-$\sigma$ uncertainty range of $\Gamma=2.475\sim2.495$ when $kT_{\rm e}=20$\,keV. In comparison, $\Gamma=2.48\sim2.51$ when $kT_{\rm e}=100$\,keV. Although a slightly harder continuum is suggested when $kT_{\rm e}$ is low, measurements of $\Gamma$ are consistent within a 3-$\sigma$ uncertainty range for different values of $kT_{\rm e}$. Similar conclusions are achieved for other sources.}

Furthermore, we run Markov chain Monte Carlo (MCMC) analysis in addition to the $\chi^{2}$ fit-goodness analysis in XSPEC in order to check any possible parameter degeneracy in our reflection model. Details can be found in Appendix\,\ref{mcmc}. The MCMC results are consistent with the uncertainty measurements given by the ERROR command in XSPEC. 

We discuss the results of our X-ray spectral analysis as following: 
\begin{itemize}
    \item The BH spin parameter $a_*$ is not well constrained in all five sources due to the lack of a clear view of the iron band in our spectra. The tightest constraint of $a_{*}$ is for \srcc\, ($a_{*}>0.88$). However, our analysis shows that all five sources are statistically consistent with a rapidly spinning BH, e.g. $a_{*}=0.9$.  Similarly, the disc inclination is not well constrained neither by the data. Most sources have an inclination angle that is consistent with either a low value ($i\approx30^{\circ}$) or a high value ($i\approx60^{\circ}$) within a $3\sigma$ uncertainty range. \srcb\ is the only case where our reflection model indicates an edge-on accretion disc with $i>70^{\circ}$.
    \item All five sources show a very high reflection fraction with $f_{\rm refl} \geq 1$, which suggests that the reflection component makes a significant contribution to the X-ray flux. No super-solar iron abundance is found. It is also interesting to note that all of our sources show a higher disc ionisation state than a typical Sy1 AGN \citep[e.g. $\log(\xi)=1-2$,][]{walton13}. We only obtain an upper limit of the disc density parameter for \srca, \srcc\, and \srcd, which are all consistent with $n_{\rm e}=10^{15}$\,cm$^{-3}$. \srcb\ and \srce\ are found to have a moderate disc density of $n_{\rm e}=10^{16}-10^{18}$\,cm$^{-3}$.
    \item The coronal emission of all five sources has $\Gamma \gtrsim 2.5$, \blue{which is softer than the continuum emission in a typical Sy1}. Such soft coronal emission is only seen in the highest flux state of some other NLS1s \citep[][]{dauser12,jiang18}.
\end{itemize}
  
\begin{table*}
    \caption{Best-fit parameters for all the sources. $F_{\rm refl}$ and $F_{\rm pl}$ are the \blue{fluxes} of the best-fit reflection and Comptonisation models in the 0.5--10\,keV band. The reflection fraction $f_{\rm refl}$ is defined as $F_{\rm refl}/F_{\rm pl}$. $F_{\rm 0.5-2keV}$ and $F_{\rm 2-10keV}$ is the absorption-corrected X-ray flux in the 0.5--2\,keV and 2--10\,keV bands respectively.}
    \label{tab_fit}
    \centering
    \begin{tabular}{cccccccc}
     \hline\hline
     Model & Parameter & Unit & \srca & \srcb & \srcc & \srcd & \srce\\
     \hline
     \texttt{relconv} & q   &     & >4 & $2.8^{+1.5}_{-1.2}$ & $4^{+3}_{-2}$ & >5 & $4^{+4}_{-2}$\\
                      & $i$ & deg & $35^{+12}_{-7}$ & $>70$ & $30^{+15}_{-7}$ & $60^{+12}_{-22}$ & $42^{+12}_{-4}$\\
                      & $a_*$ &   & $0.4^{+0.5}_{-0.9}$ & $>0.8$ & >0.88 & $>0.5$ & unconstrained \\
     \hline
     \texttt{reflionx}& $Z_{\rm Fe}$ &  $Z_{\odot}$ & $2\pm0.5$ & $1.0\pm0.2$ & $1.3\pm0.2$ & $2.0\pm0.2$ & $1.7^{+0.3}_{-0.2}$\\
                      & $\log(\xi)$ & log(erg\,cm\,s$^{-1}$) & $3.18^{+0.12}_{-0.10}$ & $3.01^{+0.08}_{-0.10}$ & $3.19^{+0.04}_{-0.15}$ & $3.3^{+0.10}_{-0.07}$ & $2.6^{+0.22}_{-0.12}$\\
                      & $\log(n_{\rm e})$ & log(cm$^{-3}$) & $<15.4$ & $17.5\pm0.2$ & <15.6 & <16.2 & $16.6^{+0.5}_{-0.4}$\\
                      & $\log(F_{\rm refl})$ & log(erg\,cm$^{-2}$\,s$^{-1}$) & $-11.72^{+0.09}_{-0.07}$ & $-11.65\pm0.03$ & $-11.68\pm0.03$ & $-11.70\pm0.06$ & $-11.957^{+0.017}_{-0.018}$\\
     \hline
     \texttt{nthcomp} & $\Gamma$ & & $2.50\pm0.02$ & $2.521^{+0.020}_{-0.012}$ & $2.57\pm0.02$ & $2.490^{+0.003}_{-0.002}$ & $2.48\pm0.02$\\
                      & $\log(F_{\rm pl})$ & log(erg\,cm$^{-2}$\,s$^{-1}$) & $-11.74\pm0.02$ & $-11.65\pm0.03$ & $-12.15\pm0.08$ & $-11.68\pm0.04$ & $-12.26^{+0.07}_{-0.08}$\\

     \hline          
     \texttt{constant} & MOS1 & & 1 & 1 & 1 & 1 & 1\\
                       & MOS2 & & $1.000\pm0.013$ & $1.006\pm0.011$ & $1.014\pm0.011$ & $0.997^{+0.019}_{-0.013}$ & $1.00\pm0.03$\\
                       & pn   & & $0.949\pm0.010$ & $0.975\pm0.008$ & $0.997\pm0.009$  & $0.972^{+0.013}_{-0.007}$ & $0.78\pm0.03$\\
     \hline
                      & $f_{\rm refl}$ & & 1.1 & 4.5 & 3.0 & 1.0 & 2.0\\
                      & $F_{\rm 0.5-2keV}$ & $10^{-12}$erg\,cm$^{-2}$\,s$^{-1}$ & 2.69 & 2.32 & 2.33 & 2.77 & 1.33\\
                      & $F_{\rm 2-10keV}$ & $10^{-12}$erg\,cm$^{-2}$\,s$^{-1}$ & 1.01 & 0.48 & 0.74 & 1.11 & 0.34\\
                      & $\chi^{2}/\nu$ &  & 345.91/297 & 471.20/380 & 337.58/325 & 453.22/337 & 174.62/148 \\
    \hline\hline 
    \end{tabular}

\end{table*}

% \begin{table}
%     \centering
%     \begin{tabular}{cccccccc}
%     \hline\hline
%         $L/L_{\rm Edd}$ & $\epsilon$ & $M_{\rm BH}/M_{\odot}$ & $A$ & $kT_{\rm db}$ (eV)\\
%         \hline
%          10 & 20\% & $1\times10^{6}$ & 10 & 22 \\
%          10 & 20\% & $1\times10^{6}$ & 300 & 4 \\
%          5 & 10\% & $1\times10^{7}$ & 10 & 15 \\
%          5 & 10\% & $1\times10^{7}$ & 300 & 2 \\
%          5 & 5\% & $1\times10^{6}$ & 10 & 9 \\
%          5 & 5\% & $1\times10^{6}$ & 300 & 3 \\
%          \hline\hline
%     \end{tabular}
%     \caption{The effective radiation temperatures given by the critical accretion disc solution in \citet[][Eq. 7.3]{shakura73}, assuming $\alpha=0.1$. The Eddington mass accretion rate is defined as $\dot{m}_{\rm Edd}c^{2}=L_{\rm Edd}/\epsilon$, where $\epsilon$ is the accretion efficiency. For super-Eddington accretion with $L<500L_{\rm Edd}$, MHD simulations estimate an accretion efficiency between $\epsilon=5\%-10\%$ \citep{jiang17}. $A$ is the ratio between the Compton energy loss and the Bremsstrahlung loss in the disc. The value of $A$ is between 10 and 300 \citep{shakura73}.}
%     \label{tab_ss73_kt}
% \end{table}

\begin{figure*}
    \centering
    \includegraphics[width=16cm]{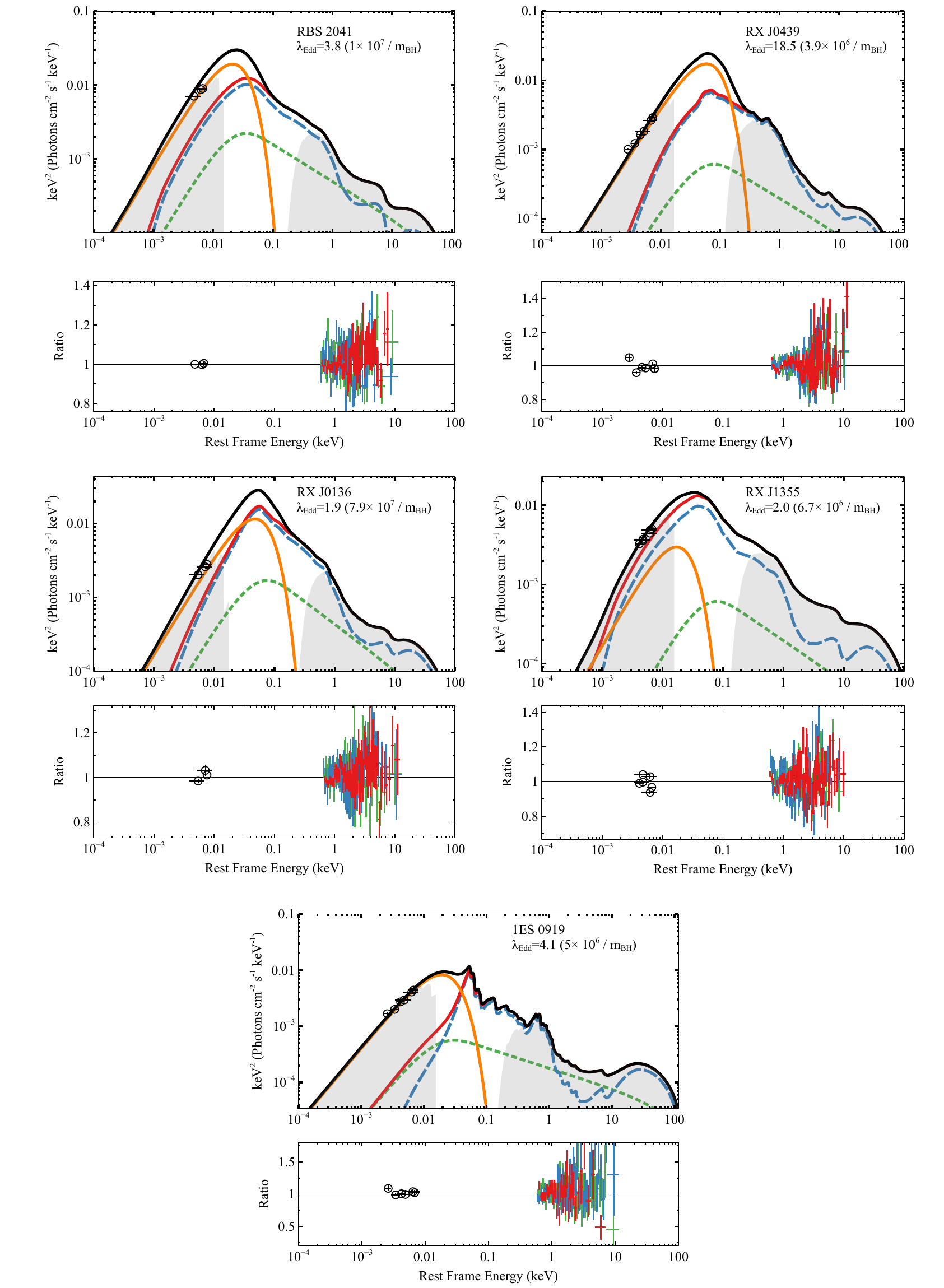}
    \caption{\red{Top panels: SED models for all five ultra-soft NLS1s. Gray shaded regions: the best-fit models; \blue{black solid lines: the best-fit models after removing Galactic absorption and extinction; red solid lines: best-fit non-thermal components after removing Galactic absorption and extinction, including relativistic disc reflection components (blue dashed lines) and thermal Comptonisation components (green dashed lines); orange solid lines:} disc thermal spectra. Bottom panels: corresponding data/model ratio plots. Red: pn; blue: MOS1; green: MOS2; black circles: OM. The error bars of OM points are smaller than the sizes of the points.}}
    \label{pic_sed}
\end{figure*}

\section{Spectral Energy Distribution} \label{sed}

So far we have obtained the best-fit model for the X-ray data, which include coronal emission and disc reflection. We extend our X-ray model to the optical and UV band.

%, assuming an inner disc temperature of $kT_{\rm db}=10$\,eV in the \texttt{reflionx} and the \texttt{nthcomp} models. 
% A disc-blackbody temperature of $kT_{\rm db}=10$\,eV is chosen according to the critical accretion disc solution in \citet{shakura73}, where the effective disc temperature is decided by the Eddington ratio $\lambda$, the accretion efficiency $\epsilon$, the BH mass $M_{\rm BH}$, and the ratio between the Compton energy loss and the Bremsstrahlung loss $A$ in the disc. The last parameter $A$ can be in the range of 10--300 \citep{shakura73}. As examples, we show the disc temperatures by given different combinations of these parameters in Table\,\ref{tab_ss73_kt}, from which we conclude that a disc blackbody of $kT_{\rm db}=10$\,eV is a good approximation in the standard disc model. Note that we are unable to measure the true temperature profile of the disc because of the Galactic absorption. 

The thermal emission from the disc is consistently modelled by the disc-blackbody model \texttt{diskbb}. \blue{The $kT_{\rm db}$ parameter of the \texttt{diskbb} model is linked to the corresponding parameters in \texttt{reflionx} and \texttt{nthcomp}.} The multiplicative model \texttt{zmshift} is applied to \texttt{diskbb} to account for the redshift. The full model is \texttt{constant * zdust * tbnew * (zmshift*diskbb + cflux*relconv*reflionx + cflux*nthcomp)} in the XSPEC format.

Note that we do not have the information about the host galaxies, such as the intrinsic dust extinction and the contribution of the star formation in the galaxies to the observed optical flux. But the combination of thermal (disc emission) and non-thermal (coronal emission and disc reflection) components can mostly describe the data very well. \blue{The best-fit disc inner temperatures $kT_{\rm db}$ and the normalisation parameters of \texttt{diskbb} for all sources are shown in Table\,\ref{tab_flux}. The other parameters listed in Table \ref{tab_fit} are also allowed to vary during our SED modelling. These parameters show consistent values as in Table\,\ref{tab_flux} and are sensitive to X-ray data only.} 

The best-fit SED models are shown in Fig.\,\ref{pic_sed}, and the flux of the thermal and non-thermal components are shown in \third{Table} \,\ref{tab_flux}. $F_{\rm th}$ and $F_{\rm non-th}$ are the flux of the thermal and non-thermal components respectively in the 0.1\,eV--100\,keV band given by our best-fit SED models. $f_{\rm non-th}$ is the flux ratio of the non-thermal emission and the total emission in the same energy band. Similarly, $F^{\rm opt}_{\rm th}$, $F^{\rm opt}_{\rm non-th}$ and $f^{\rm opt}_{\rm non-th}$ are calculated in the 1--10\,eV band (optical and UV). The corresponding Eddington ratio $\lambda=4\pi D^{2} (F_{\rm th}+F_{\rm non-th})/L_{\rm Edd}$ is labelled in each panel of Fig.\ref{pic_sed} where $L_{\rm Edd}$ is calculated using the BH masses given in Table\,\ref{tab_obs}.

\red{Assuming the BH mass measurements are all accurate, \srcb\ shows the highest Eddington ratio, which is approximately 19 times the Eddington limit, and \srcc\ shows the lowest Eddington ratio, which is around the Eddington limit. Our inferred Eddington ratios are similar to the values obtained by the warm corona models. For instance, \citet{jin09} found that the warm corona model suggests an Eddington ratio of $\lambda_{\rm Edd}\approx2.7$ after adopting the same BH mass for \srcc\ as we do. However, our model predicts a higher fraction of disc thermal emission in the UV band. Readers may compare Fig.\,\ref{pic_sed} with Fig.\,4 in \citet{jin09}.}

Note that \srcc\ and \srcb\ have a similar luminosity of $1\sim2\times10^{46}$\,erg\,s$^{-1}$. The difference of their inferred Eddington ratios in our work is because \srcc\ was estimated to have a BH mass more than one order of magnitude higher than \srcb\ \citep[e.g.][]{jin09,grupe10}. \red{However, the systematic uncertainty of the BH mass measurements using $H_{\beta}$ line width is very large, depending on the assumption for the geometry of the BLR \citep[e.g. $\Delta(\log(m_{\rm BH}))=0.5$,][]{kaspi00, mclure04}. Moreover, the correction for radiation pressure onto the BLR may add more uncertainty to the mass measurements \citep{marconi08}. Therefore, they may share a similar Eddington ratio if \srcc\ and \srcb\ have a similar true BH mass. Nevertheless, we conclude that the five ultra-soft NLS1s in our sample share a similar bolometric luminosity and show an accretion rate around or a few times the Eddington limit.}

\begin{table*}
    \centering
    \begin{tabular}{ccccccccccc}
    \hline\hline
      Source & $kT$ (eV) & norm & $F_{\rm tol}$ & $F_{\rm non-th}$ & $f_{\rm non-th}$ & $F^{\rm opt}_{\rm tol}$ & $F^{\rm opt}_{\rm non-th}$ &$f^{\rm opt}_{\rm non-th}$ & $\lambda_{\rm Edd}$ &$\chi^{2}/\nu$ \\
    \hline
      \srca  & $9\pm1$ & $(4.3\pm0.7)\times10^8$ & 10.7 & 5.2 & 49\% & 2.5 & 0.7 & 28\% & 3.8 & 353.03/298 \\
      \srcb  & $19\pm1$ & $(1.3\pm0.2)\times10^7$ & 5.7 & 2.1 & 37\% & 0.5 & 0.07 & 14\% & 18.5 & 539.61/385\\
      \srcc  & $6.20\pm0.09$ & $(1.1\pm0.2)\times10^{7}$ & 7.6 & 4.8 & 62\% & 0.6 & 0.2 & 33\% & 1.9 & 404.32/326\\
      \srcd  & $2^{+4}_{-1}$ & $(2.2\pm0.2)\times10^{10}$ & 4.6 & 4.2 & 90\% & 1.1 & 0.7 & 64\% & 2.0 & 531.04/341\\
      \srce  & $7^{+2}_{-1}$ & $(4.4\pm0.1)\times10^8$ & 4.0 & 1.7 & 43\% & 0.9 & 0.06 & 7\% & 4.1 & 263.45/152\\
     \hline\hline
    \end{tabular}
        \caption{\red{The best-fit parameters of the \texttt{diskbb} model given by SED modelling, and \blue{the flux of the thermal (disc emission) and non-thermal (coronal emission and disc reflection) components} inferred by our model. $F_{\rm tol}$ and $F_{\rm non-th}$ are the flux of the total emission and the non-thermal emission respectively calculated in the 0.1\,eV--100\,keV band. $F^{\rm opt}_{\rm tol}$ and $F^{\rm opt}_{\rm non-th}$ are the flux of the same components calculated in the 1\,eV--10\,eV band. All the flux values are in the units of $10^{-11}$\,erg\,cm$^{-2}$\,s$^{-1}$. $f_{\rm non-th}$ and $f^{\rm opt}_{\rm non-th}$ are the percentages of non-thermal emission in the 0.1\,eV--100\,keV and 1--10~eV bands respectively. $\lambda_{\rm Edd}$ is the Eddington ratio estimated by using $F_{\rm tol}$ and assuming $m_{\rm BH}$ in Table\,\ref{tab_obs}.}}
    \label{tab_flux}
\end{table*}

It is interesting to mention that our best-fit SED models suggest the non-thermal emission, including coronal emission and disc reflection, is responsible for more than 50\% of the total flux of \srcb, \srcc\, and \srcd\, in the 0.1\,eV--100\,keV band. The most extreme case is \srcd, where the inferred non-thermal emission fraction is around 90\% in the full band and 64\% in the 1--10\,eV (optical) band. The high fraction of non-thermal emission at longer wavelengths is due to the extremely soft coronal emission and the reflection from a highly ionised inner disc region. 
%In comparison, we refer interested readers to \citet{grupe10} for an sample of SED models for sources with a much harder X-ray continuum ($\Gamma < 2$), where the UV and optical emission is dominated by the thermal emission from the disc. 

Our results suggest that a significant fraction of disc energy in ultra-soft NLS1s is not radiated away from the disc surface as in the \citet{shakura73}, but transferred to the coronal region and carried away in the form of non-thermal emission \citep[][]{haardt91,svensson94}.

\section{Conclusions} \label{conclude}

\red{We analyse the \xmm\ observations of five ultra-soft NLS1s using a relativistic disc reflection model in this work. A reflection-based SED model is able to describe the simultaneous OM and EPIC observations very well. Our reflection models indicate a more ionised disc in ultra-soft NLS1s compared to other typical Sy1s. The best-fit SED models suggest that these sources share a similar luminosity, corresponding to an Eddington ratio of $\lambda_{\rm Edd}=1-20$ assuming previous BH mass measurements.} In particular, our models suggest that a significant fraction of the disc energy is carried away in the form of non-thermal emission instead of thermal emission from the surface of the disc. In the most extreme case, the optical emission of \srcd\ is dominated by non-thermal emission from the innermost region.

\red{As explained above, the S/N of the \xmm\ data in the iron band is not high enough to enable us to constrain the iron line profile due to the nature of these ultra-soft NLS1s: the broad band spectral analysis by using a reflection model suggests that the disc density is low ($n_{\rm e}<10^{18}$\,cm$^{-3}$) and the ionisation state is particularly high. At a high ionisation state (e.g. $\log(\xi)>3$), the surface of the disc becomes so ionised that the emission and absorption features in the reflection spectrum become very weak \citep{ross93,garcia10}. Additionally, their ultra-soft X-ray continuum emission of $\Gamma\approx2.5$ makes their iron band observation particularly challenging.} 

\begin{figure}
    \centering
    \includegraphics{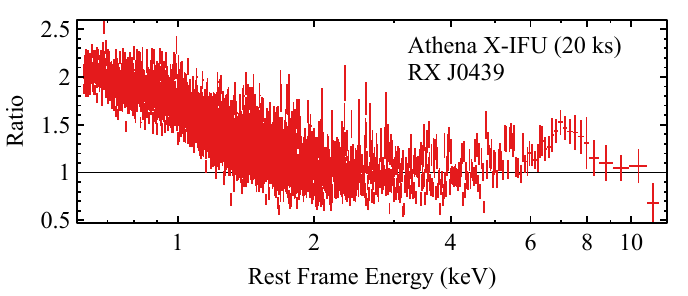}
    \caption{\red{The data/model ratio plot for a simulated \textit{Athena} X-IFU spectrum of \srcb\ using an absorbed power-law model. The simulations assume a net exposure of only 20\,ks and are calculated using the best-fit disc reflection model presented in Fig.\,\ref{pic_fit}. The spectrum has been grouped significantly for demonstration purposes.}}
    \label{pic_athena}
\end{figure}

\red{Future observations with higher S/N and energy resolutions in the iron band, e.g. from \textit{Athena}, will be able to better constrain the spectral shape in the iron band and thus distinguish the reflection interpretation from other models, such as warm corona \citep[e.g.][]{porquet19,ballantyne20}. As an example, we present a simulated \textit{Athena} IFU spectrum of \srcb\ with a net exposure of 20\,ks in Fig.\,\ref{pic_athena}. The spectrum is calculated using the best-fit model obtained in Section\,\ref{xray}. According to our simulations, \textit{Athena} will be able to detect not only strong soft excess emission but also clear evidence for a broad Fe K emission line assuming the right reflection model.}

% We also try the warm corona model for \srccfull, the most extreme source in our sample. Previous analyses suggested an inner disc temperature of $kT_{\rm db}=7.9\pm0.3$\,eV when using a warm corona-based SED model \citep{jin09}. However, blackbody seed photons were assumed in their warm corona model while the disc emission was modelled by a disc-blackbody model. After correcting for the seed photon spectrum, we find that an inner disc temperature of $kT_{\rm db}>20$\,eV is required to account for the observed UV flux in the simultaneous OM observation of \srccfull. Such a high temperature is inconsistent with the previous results. The true temperature of the disc in \srccfull\ is unknown due to the Galactic absorption. We only point out the inconsistency in the current warm corona model for this source.

\section*{Acknowledgements}

J.J. acknowledges support by the Cambridge Trust and the Chinese Scholarship Council Joint Scholarship Programme (201604100032), the Tsinghua Astrophysics Outstanding (TAO) Fellowship, the Tsinghua Shuimu Scholar Programme. A.C.F. acknowledges support by the ERC Advanced Grant 340442. M.L.P. is supported by European Space Agency (ESA) Research Fellowships.

\section*{DATA AVAILABILITY}

The data underlying this article are available in the High Energy Astrophysics Science Archive Research Center (HEASARC), at https://heasarc.gsfc.nasa.gov.
%%%%%%%%%%%%%%%%%%%%%%%%%%%%%%%%%%%%%%%%%%%%%%%%%%



\bibliographystyle{mnras}
\bibliography{softagn.bib} % if your bibtex file is called example.bib

%%%%%%%%%%%%%%%%%%%%%%%%%%%%%%%%%%%%%%%%%%%%%%%%%%

%%%%%%%%%%%%%%%%% APPENDICES %%%%%%%%%%%%%%%%%%%%%

\appendix

\section{MCMC Analysis} \label{mcmc}

We check the constraints of all the parameters in the reflection model by using the MCMC algorithm. The XSPEC/EMCEE code by Jeremy Sanders based on the python implementation  \citep{foreman12} and the MCMC ensemble sampler \citep{goodman10} was used. We use 50 walkers with a length of 250000, burning the first 5000. A convergence test has been conducted and the Gelman-Rubin scale-reduction factor $R<1.3$ for every parameter. Fig.\,\ref{pic_mcmc_rbs} to \ref{pic_mcmc_1es} show the output distributions of all the parameters. We do not find obvious evidence for parameter degeneracy. The uncertainty ranges of parameters given by our MCMC analysis are consistent with the measurements using the ERROR command in XSPEC.

\begin{landscape}
\begin{figure*}
    \centering
    \includegraphics[width=21cm]{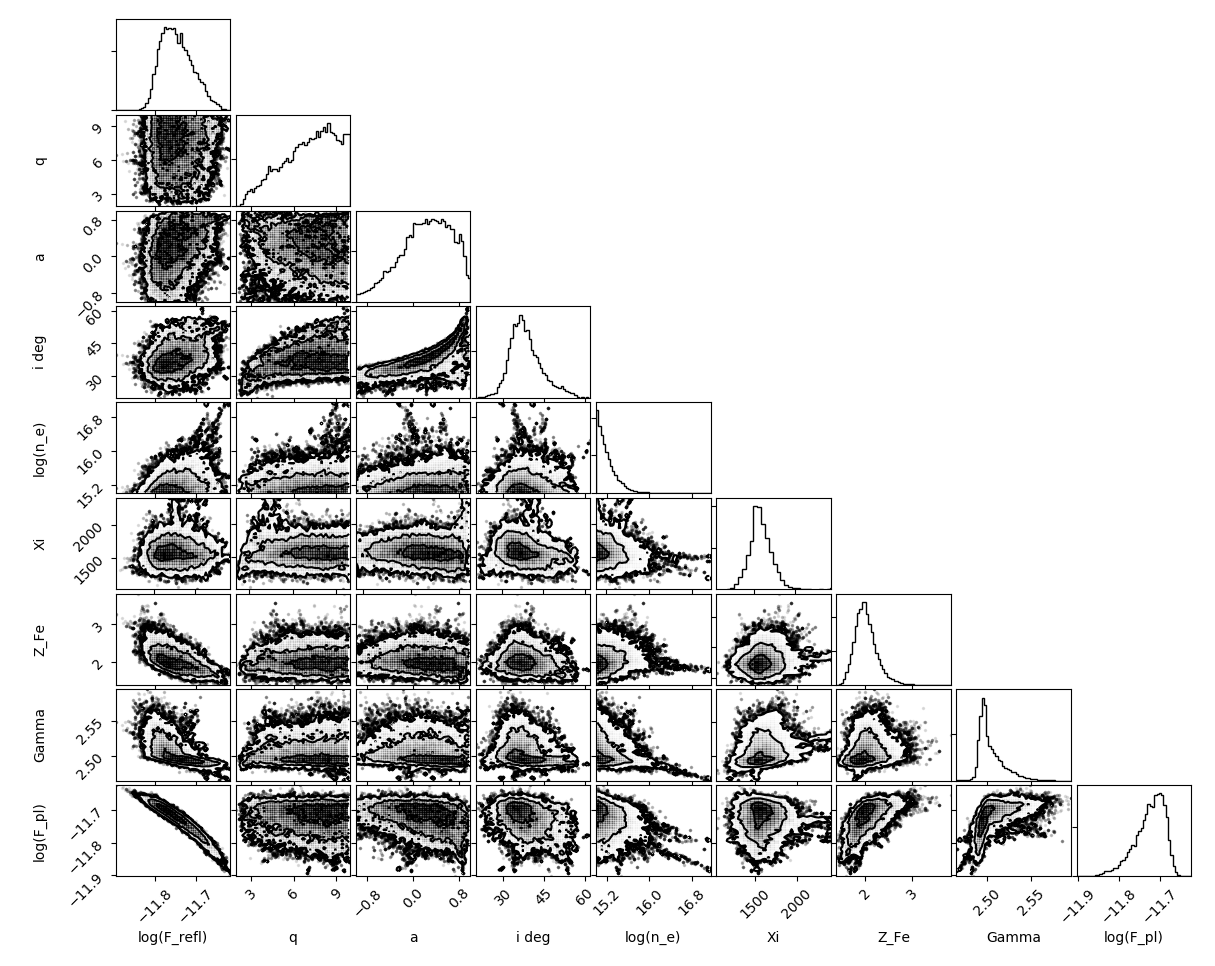}
    \caption{Output  distributions  for  the  MCMC  analysis  of  the  best-fit  models  of  the EPIC  spectra  of  \srca.  Contours correspond to 1, 2 and 3\,$\sigma$. All the parameters are in the same units as in Table\,\ref{tab_fit}.}
    \label{pic_mcmc_rbs}
\end{figure*}

\begin{figure*}
    \centering
    \includegraphics[width=21cm]{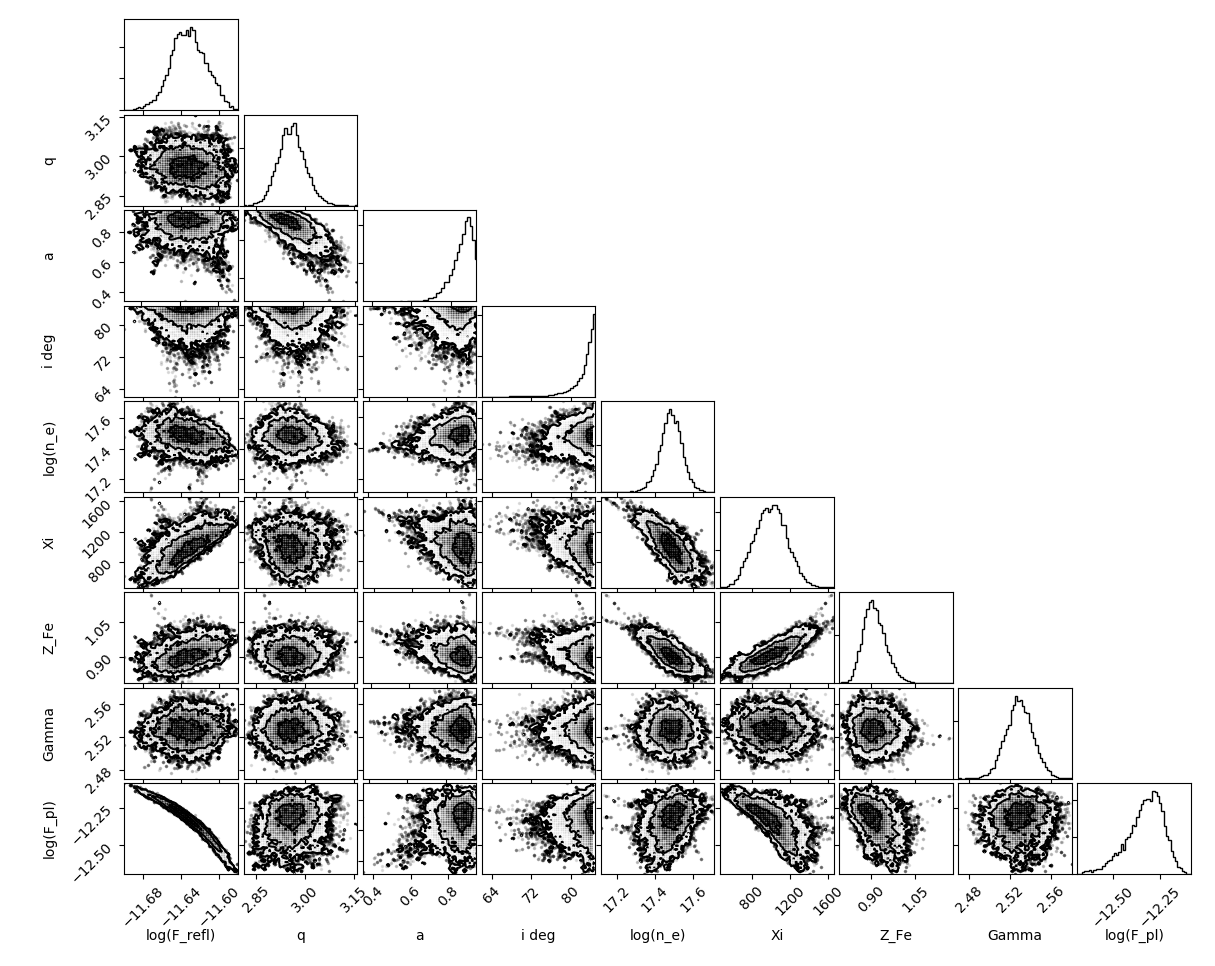}
    \caption{Output  distributions  for  the  MCMC  analysis  of  the  best-fit  models  of  the EPIC  spectra  of  \srcb.}
    \label{pic_mcmc_rxj0439}
\end{figure*}

\begin{figure*}
    \centering
    \includegraphics[width=21cm]{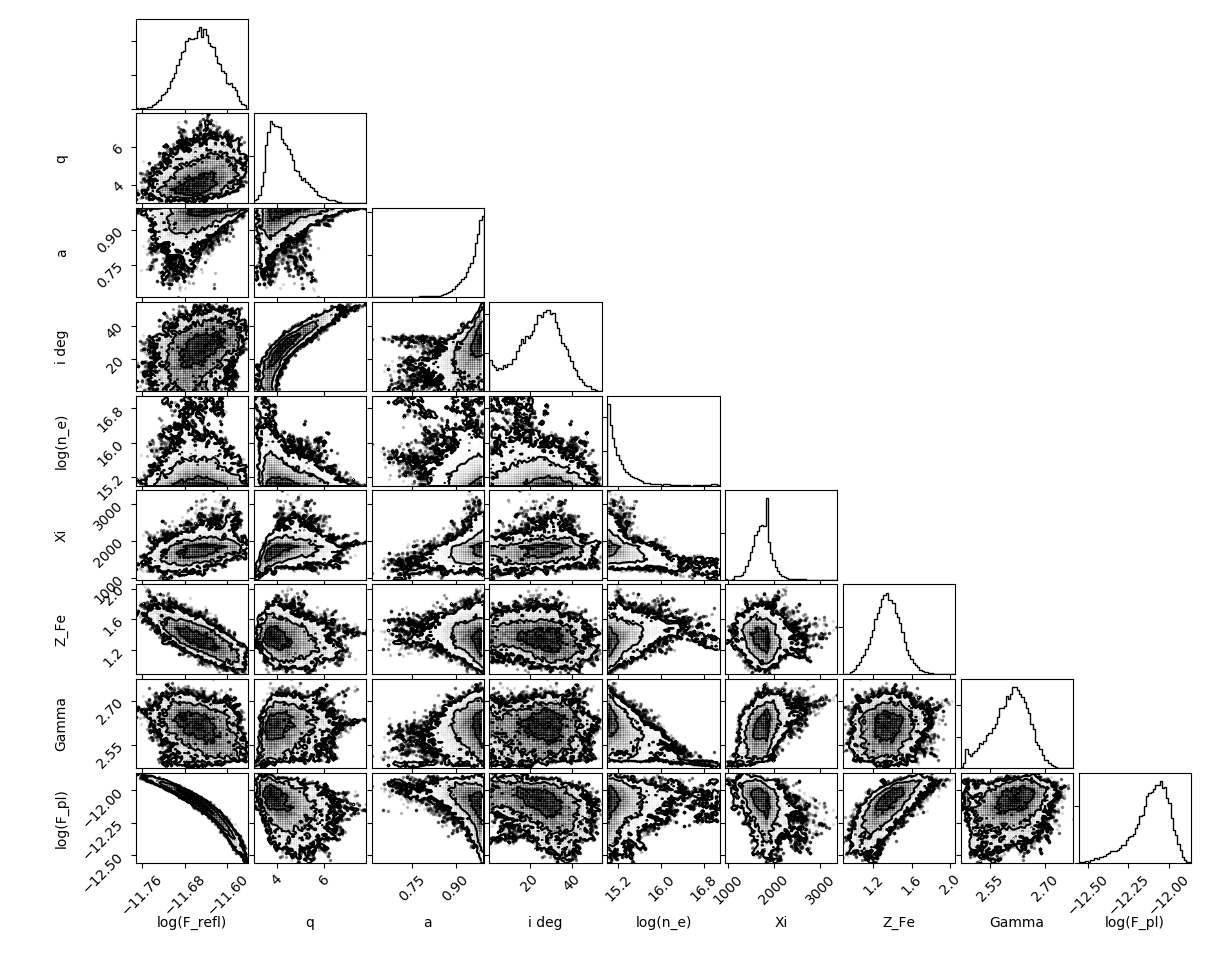}
    \caption{Output  distributions  for  the  MCMC  analysis  of  the  best-fit  models  of  the EPIC  spectra  of  \srcc.}
    \label{pic_mcmc_rxj0136}
\end{figure*}

\begin{figure*}
    \centering
    \includegraphics[width=21cm]{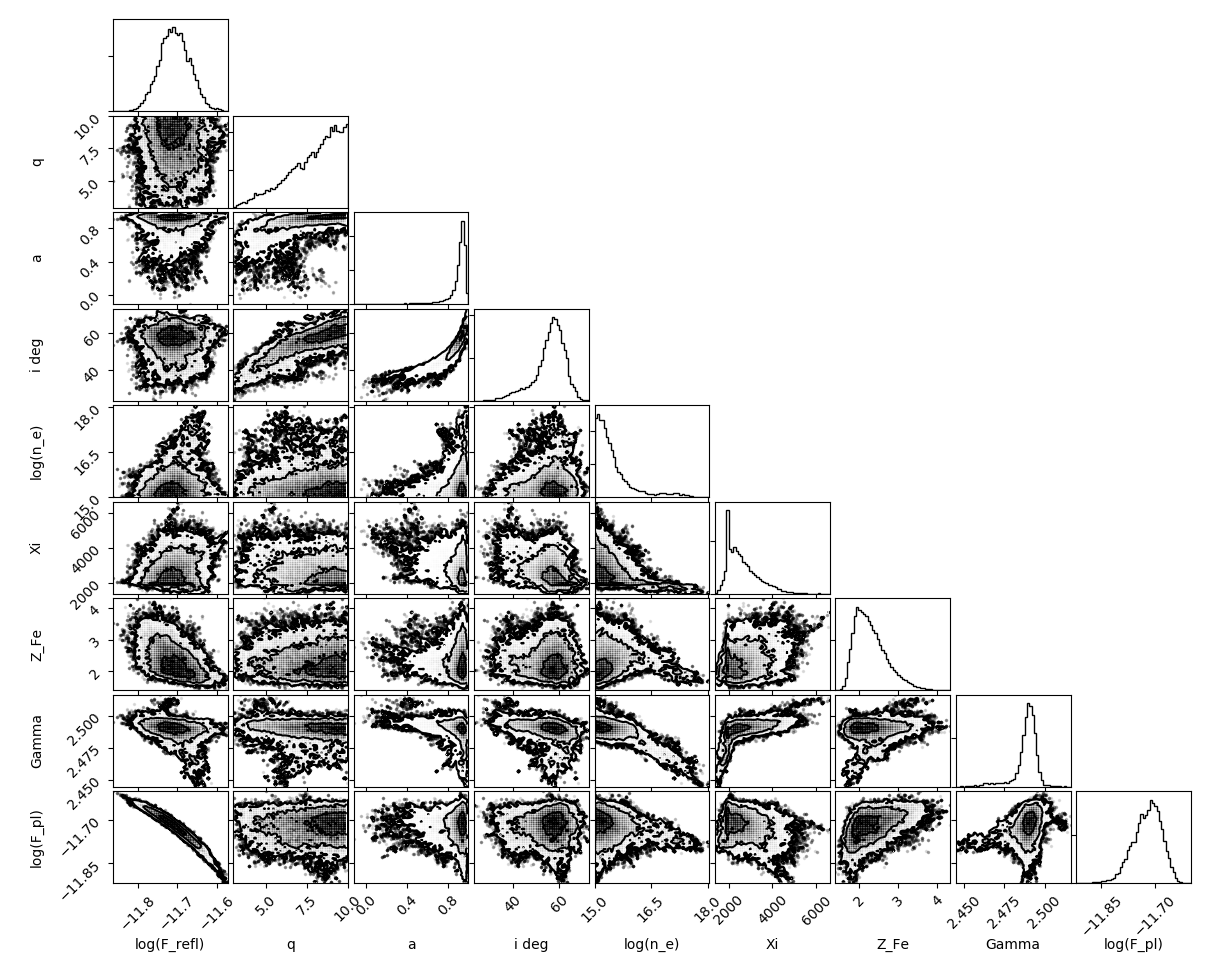}
    \caption{Output  distributions  for  the  MCMC  analysis  of  the  best-fit  models  of  the EPIC  spectra  of  \srcd.}
    \label{pic_mcmc_rxj1355}
\end{figure*}

\begin{figure*}
    \centering
    \includegraphics[width=21cm]{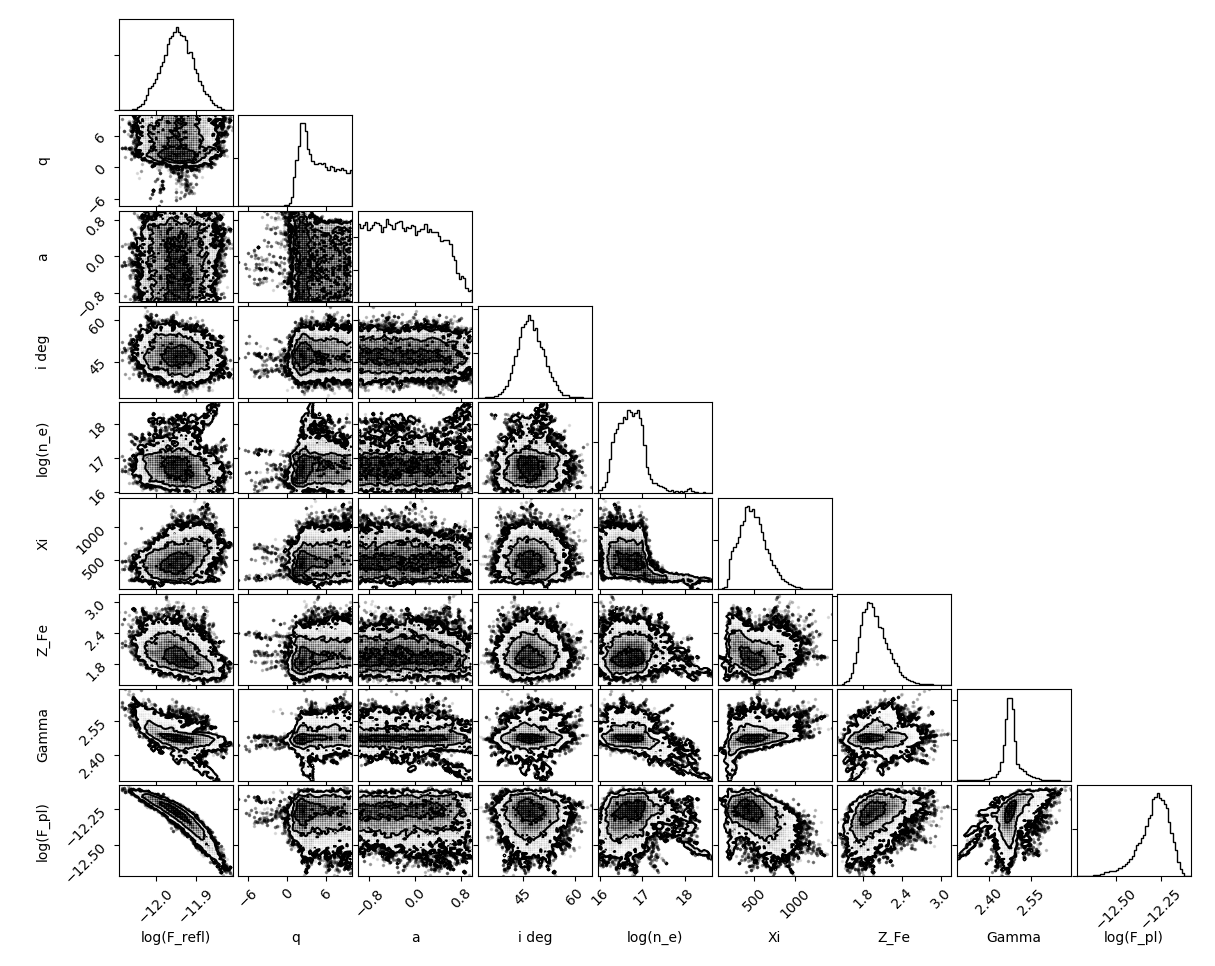}
    \caption{Output  distributions  for  the  MCMC  analysis  of  the  best-fit  models  of  the EPIC  spectra  of  \srce.}
    \label{pic_mcmc_1es}
\end{figure*}

\end{landscape}

%%%%%%%%%%%%%%%%%%%%%%%%%%%%%%%%%%%%%%%%%%%%%%%%%%

% Don't change these lines
\bsp	% typesetting comment
\label{lastpage}
\end{document}